\documentclass[preprint, double-spaced]{revtex4}
\usepackage{graphicx}
\usepackage{amsmath,bbm}
\usepackage{amssymb,bm}

\begin{document}

\author{C.~P.~Singh, P.~K.~Srivastava, S.~K.~Tiwari}
\affiliation{Department of Physics, Banaras Hindu University, 
Varanasi 221005, INDIA}

\title{QCD Phase Boundary and Critical Point in a Bag Model Calculation}

\begin{abstract}
 Location of critical point and mapping the QCD phase boundary still exists as one of the most interesting and studied problems of heavy-ion physics. A new equation of state(EOS) for a gas of extended baryons and pointlike mesons is presented here which accounts for the repulsive hard-core interactions arising due to the geometrical size of the baryons. A first order deconfining phase transition is obtained using Gibbs' equilibrium criteria and a bag model EOS for the weakly interacting quark matter. It is interesting to find that the phase transition line ends at a critical point beyond which a cross-over region exists between hot-dense meson gas and quark-antiquark gluon matter. Our curve resembles in shape closely with the predicitions of the available lattice gauge calculations and also reproduces the conjectured phase boundary.

\end{abstract}
\maketitle

The existence of critical point in the studies of QCD phase diagram has attracted considerable theoretical and experimental attention recently. The phase diagram of quark matter is still not understood either experimentally or theoretically. The conjectured phase boundary between quark gluon plasma (QGP) and hot, dense hadron gas (HG) represents a first-order phase transition line for nonzero and moderate values of temperature T and baryon chemical potential ($\mu_{B}$) [1-3]. At extremely high baryon densities (i.e., large $\mu_{B}$), we expect a colour-flavour-locked (CFL) phase involving colour-superconducting quark matter. As we increase T and decrease $\mu_{B}$, it is also expected that the first-order phase transition line ends at a critical point beyond which there exists a cross-over region since thermal fluctations at temperature (say $T > 170$ MeV) break up mesons ( mostly pions) which are densly populated in this region and this thus results into a gas of quarks, antiquarks, gluons etc. The existence of such a critical point was proposed a long time ago [4,5] and more recently its properties were investigated in detail with the help of several models [6,7]. We are hopeful that the experiments with different colliding beam energies at Relatvistic Heavy-Ion Collider (RHIC) will provide a suitable experimental window [8] for the search of QCD critical point as well as for mapping the QCD phase boundary. Indeed we have gained reliable insight into the thermodynamics of QGP from lattice QCD calculation and our knowledge about its dynamics is particularly helpful in the high temperature limit where it becomes weakly coupled, However, RHIC has given us result that, at least, at temperatures within a factor of two of that at which hadrons melt, the dynamics of QGP is closer to an ideal liquid limit rather than to the ideal gas limit indicating the presence of a strongly coupled QGP. Confusions still prevail about the theoretical understanding of the QCD phase transition. We do not know whether the conjectured phase boundary is an outcome of deconfinement and/or chiral symmetry restoration. The purpose of this paper is to determine the phase boundary and to locate the critical point in a first order deconfining phase transition obtained by using EOS for the interacting quark matter and HG, separately.

 We proposed a new thermodynamically consistent EOS for the HG where the geometrical size of the baryons is explicitly incorporated as the excluded volume correction and our model uses full quantum statistics in the partition function of the grand canonical ensemble so that there arises no problem in dealing with large $\mu_{B}$  and low T region and thus the full phase boundary in T, $\mu_{B}$ plane can be investigated easily. In the earlier version of the model [9,10], we have simplified the calculations by using Boltzmann approximation and we have noticed that the model successfully describes the observed particle yields, particle ratios etc at the chemical freezeout of the HG fireball in the heavy-ion collisions [10]. In order to determine the thermodynamic properties of the weakly interacting quark matter, we use a simple bag model EOS with the perturbative corrections of the order of $\alpha_S^{3/2}$ in strong interaction coupling constant $\alpha_{S}$. The advantage of the bag model clearly lies in determining the thermodynamic parameters in the region of nonzero as well as large baryon chemical potential $\mu_{B}$ which is still not properly accessible in the lattice calculations. We obtain the full phase boundary by Gibbs' construction of equlibrium phase transition between QGP and HG. We find a significant result that the first-order phase transition line ends at a point beyond which there occurs no phase transition and the cross-over region is only present. Thus we determine the precise coordinates of the QCD critical point in the phase diagram and we compare the  location with the predictions of other models including lattice calculations. The construction of the QCD phase boundary by comparing EOS of weakly interacting QGP with a bag pressure term to EOS of hadron gas with an excluded volume correction is not new and was done by several authors [11-14]. Excluded volume corrections in many of these approaches have usually been incorporated in a thermodynamically incosistent manner. Our approach has following new and interesting features :(1) Our EOS for the HG is thermodynamically consistent and we have obtained the chemical freeze out curve using the same formulation.(2) We have used quantum statistics in our formulation so that we can determine the phase boundary in the entire (T,$\mu_{B}$) plane.(3) We find that our calculated phase diagram almost reproduces the conjectured QCD phase diagram and the coordinates of the critical point matches well with the lattice prediction. No other model reproduces the features so well.(4) Most importantly, we get a first-order deconfining phase transition line where other models including lattice calculations reveal chiral phase transition.(5) The chemical freeze out curve obtained from our formulation lies in close proximity to the critical point and this supports the suggestions of previous authors [3].

Let us first consider QGP and we assume that it consists of massless quarks (u,d), their antiquarks and gluons only. So the pressure of QGP can be written as [15]:

\begin{equation}
  \begin{split}
 P_{QGP}
= \frac{37}{90}\pi^2 T^4 + \frac1{9}\mu_B^2 T^2 + \frac{\mu_B^4}{162 \pi^2} 
\\
- \alpha_S\left[\frac{11}{9}\pi T^4 + \frac{2}{9\pi^2}\mu_B^2 T^2 + \frac1{81\pi^3}\mu_B^4\right ]
\\
+\frac{8\alpha_S^{3/2} T}{3\pi^2\sqrt{2\pi}}{\left[\frac{8\pi^2 T^2}{3} +\frac2{9}\mu_B^2\right ]^{3/2}} - B
 \end{split}
\end{equation}
where $\mu_B$, T dependence of $\alpha_S$ can be given as [15]:

\begin{equation}
\alpha_S
= \frac{12\pi}{29}\left[ln  (\frac{0.089 \mu_B^2 + 15.622 T^2}{\Lambda^2})\right]^{-1}
\end{equation}
Here we have used $B^{1/4} = 216 ~MeV$ and $\Lambda = 100 ~MeV$ in our calculation.

The grand canonical partition function for the HG with full quantum statistics and after incorporating excluded-volume correction in a thermodynamically consistent manner, can be written as [16]:

\begin{equation}
\begin{split}
ln Z_i^{ex} = \frac{g_i}{6 \pi^2 T}\int_{V_i^0}^{V-\sum_{j} N_j V_j^0} dV
\\
\int_0^\infty \frac{k^4 dk}{\sqrt{k^2+m_i^2}} \frac1{[exp\left(\frac{E_i - \mu_i}{T}\right)+1]}
\end{split}
\end{equation}
where $g_i$ is the degeneracy factor of ith species of baryons, E is the energy of the particle ($E=\sqrt{k^2+m_i^2}$), $V_i^0$ is the eigen volume of one ith species baryon and $\sum_{j}N_jV_j^0$ is the total occupied volume.

We can write Eq.(3) as:

\begin{equation}
ln Z_i^{ex} = V(1-\sum_jn_j^{ex}V_j^0)I_{i}\lambda_i
\end{equation}
where
\begin{equation}
I_i=\frac{g_i}{6\pi^2 T}\int_0^\infty \frac{k^4 dk}{\sqrt{k^2+m_i^2}} \frac1{\left[exp(\frac{E_i}{T})+\lambda_i\right]}
\end{equation}
and $\lambda_i = exp(\frac{\mu_i}{T})$ is the fugacity of the particle, $n_j^{ex}$ is the number density of jth type of baryons after excluded volume correction and can be obtained from Eq.(4) as:
\begin{equation}
n_i^{ex} = \frac{\lambda_i}{V}\left(\frac{\partial{ln Z_i^{ex}}}{\partial{\lambda_i}}\right)_{T,V}
\end{equation}
This leads to a transcedental equation as
\begin{equation}
n_i^{ex} = (1-R)I_i\lambda_i-I_i\lambda_i^2\frac{\partial{R}}{\partial{\lambda_i}}+\lambda_i^2(1-R)I_i^{'}
\end{equation}
where
\begin{equation}
I_i^{'}=\frac{\partial{I_i}}{\partial{\lambda_i}}=-\frac{g_i}{6\pi^2 T}\int_0^\infty \frac{k^4 dk}{\sqrt{k^2+m_i^2}} \frac1{\left[exp(\frac{E_i}{T})+\lambda_i\right]^2}
\end{equation}
and $R=\sum_in_i^{ex}V_i^0$ is the fractional occupied volume. We can write R in an operator equation:
\begin{equation}
R=\hat{R}+\Omega R
\end{equation}
where $\hat{R}=\frac{R^0}{1+R^0}$ with $R^0 = \sum n_i^0V_i^0 + \sum I_i^{'}V_i^0\lambda_i^2$; $n_i^0$ is the density of pointlike baryons of ith species and the operator $\Omega$ is:
\begin{equation}
\Omega = -\frac{1}{1+R^0}\sum_i n_i^0V_i^0\lambda_i\frac{\partial}{\partial{\lambda_i}}
\end{equation}
Using Neumann iteration method in Eq.(9) and retaining the series upto $\Omega^2$ term, we get
\begin{equation}
R=\hat{R}+\Omega \hat{R} +\Omega^2\hat{R}
\end{equation}

Eq.(11) can be solved numerically. The total pressure [15] of the hadron gas after excluded volume correction is:
\begin{equation}
P_{HG}^{ex} = T(1-R)\sum_iI_i\lambda_i + \sum_iP_i^{meson}
\end{equation}

In (12), the first term represents the pressure due to all types of baryons and the second term gives the total pressure due to all mesons having pointlike size only. This makes it clear that we consider the hard-core repulsion existing between two baryons only.

We have considered all the baryons and the mesons as well as their resonances having masses upto 2 GeV/$c^2$ in our calculation. In order to conserve strangeness quantum number, we have used the criterion of equating the net strangeness equal to zero, i.e., $\sum_i S_i(n_i^S - n_i^{\bar{S}})=0$ where $S_i$ is the strangeness quantum number of ith hadron, $n_i^S$ and $n_i^{\bar{S}}$ are the strange hadron density of ith hadron and ith anti-hadron, respectively. Strangeness neutrality condition yields the value of strange chemical potential in terms of $\mu_{B}$. We have considered mesons as pointlike particles in this calculation. Furthermore, we have taken an equal volume $V^0=\frac{4 \pi r^3}{3}$ for each type of baryon with a hard-core radius r=0.8 fm. The first-order phase transition boundary is determined by using the Gibbs' equilibrium condition $P_{HG}^{ex}(T_c,\mu_c)=P_{QGP}(T_c,\mu_c)$.

In Fig.1, we have shown the phase boundary between QGP and HG as obtained from our calculations. We start from a low but nonzero value of T and large value of $\mu_{B}$ and we move towards large T and small( and nonzero) $\mu_{B}$. We find a first order phase transition line and it ends at a QCD critical point.The coordinates are $T_{c}$=160 MeV and $\mu_{c}$=156 MeV. The critical point as obtained by us lies close to the lattice result LTE04 [17]. Since the value of $\mu_{c}$ is the lowest in comparision to all other models, we hope that this point can be reached in the RHIC experiments. From this critical point to the $\mu_{B}=0$ line , we find a transitional cross-over region which cannot be described or modelled analytically. Hadronic degrees of freedom are insufficient to give a valid description [18] of this region whereas free quark and gluons start playing a significant role only at much higher temperatures. We have compared our prediction regarding the location of the QCD critical point with those obtained from different models. We find that critical point in our curve lies closer to the lattice gauge predictions. It should be emphasized that the location of the critical point in our calculation means the end point of the first order phase transition line and beyond this point, Gibbs' equilibrium condition does not remain valid. Here we stress that the dependence of our results on the values of two parameters, the bag constant B and the hard core radius r is small. We have shown in fig.1 by the curves $P_{1}$, $P_{2}$ and $P_{3}$ respectively. We find that the location of the critical point shifts from $C_{1}$ to $C_{2}$ as we decrease the value of the bag constant B. Furthermore, the variation in the QCD scale factor $\Lambda$ do not give any substantial change.

It is very difficult to predict the coordinates of the critical point reliably and this is also evident by the plot in Fig.1 where we find that the predictions of different models vary wildly [2]. However, it has been suggested that the present heavy-ion experiments can be used to locate the QCD critical point [7]. We find that the critical point as obtained from our calculation lies in the region of the phase diagram accessible at the current energy of 
\begin{figure}
 \includegraphics[height=30em]{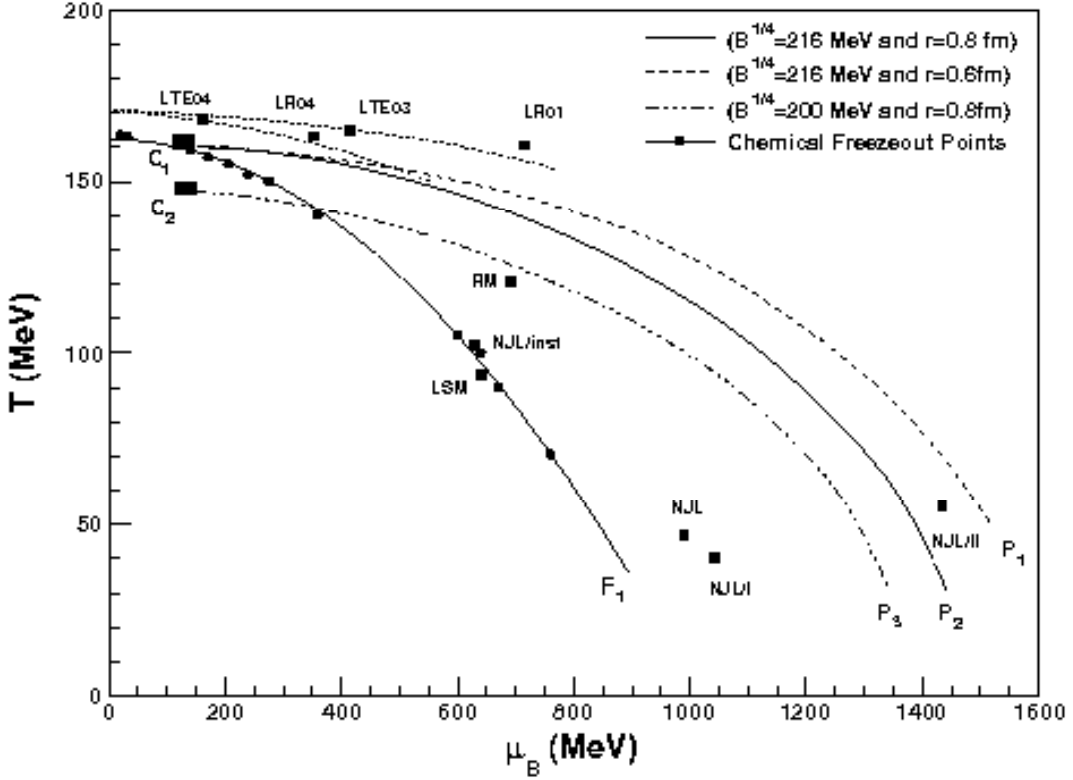}
\caption[]{The location of the QCD critical point in the QCD phase diagram as calculated in our model.$P_{2}$ is the phase boundary with $B^{1/4}=216 MeV$ and r=0.8 fm, $P_{1}$ is with $B^{1/4}=216 MeV$ and r=0.6 fm, and $P_{3}$ is with $B^{1/4}=200 MeV$ and r=0.8 fm. $F_{1}$ is the chemical freeze out line obtained in our model. $C_{1}(T_{c}=160 MeV, \mu_{c}=156 MeV)$ and $C_{2}(T_{c}=146 MeV, \mu_{c}=156 MeV)$ are the critical end points on $P_{1}, or  P_{2} and P_{3}$, respectively. Critical points denoted by LR04[24], LR01[25], LTE03[26], LTE04[17] are lattice model results and NJLinst [5],LSM,NJL [20],NJL/I,NJL/II [21],RM [22] are in other models and the points and the notations have been taken from ref. [2].}
\end{figure}200 GeV/nucleon at which RHIC explores the cross-over region. Near the critical point, chemical freezeout points are also helpful in finding its location. It has been suggested that the experimental observables should show nonmonotonic behaviour as a function of centre-of-mass energy $\sqrt{s}$ when the freeze out point lies close to the critical point [1,19]. We have determined the locations of freeze out points in various heavy-ion experiments by measuring the ratios of particle yields and fitting to our excluded volume HG model with T and $\mu_{B}$ as parameters. We have plotted in Fig.1, the freezeout curve as obtained from our model. We find that the critical point lies almost on (or near) the freeze out curve.This endorses the usefulness of the finding of Stephanov, Rajagopal and Shuryak [3] because they have shown that a non-monotonic behaviour of fluctations (eg, of multiplicity) can be considered as a signal for the critical point. However after the critical point, the difference between the freeze out curve and the phase transition line increases as $\mu_{B}$ increases and T decreases. The freeze out points tend to cluster near  the QCD critical point.

In conclusion, we have demonstrated the first order phase transition boundary in a simple bag like model describing the deconfining phase transition of quarks and gluons. It should be stressed that our excluded volume model proposed in this paper is thermodynamically consistent and also incorporates full quantum statistics. We have also determined the precise location of QCD critical point using new EOS for the HG proposed by us. The results are in agreement with what we expect from lattice calculations [17]. It should be emphasized that the lattice calculations [23-26] have failed so far to converge on a prediction for the location of the critical point. However, our interpretation differs from other QCD models which are all based on the chiral dynamics. Obviously the existence of the critical point in all these calculations follow from the basic assumption, that the finite $\mu_{B}$ chiral phase transition is first order.However, the picture based on the chiral dynamics in the baryon-dense region casts a shadow of doubt as CFL phase breaks chiral symmetry. The occurrence of a novel phase of dense quarks, named as quarkyonic phase was recently proposed based on the large $N_{c}$ argument where $N_{c}$ denotes number of colours [27]. This phase occurs above $\mu_{B}=M_{B}$ where $M_{B}$ is the baryon mass and is characterized by non-vanishing baryon number density and confinement. The clear separation of the quarkyonic phase from the hadronic phase is lost in a system with finite $N_{c}$ but any large change in the baryon number density can reveal a quarkyonic transition. Our model endorses the deconfining nature of the first order phase transition. We also find the existence of a cross-over region lying beyond the critical point where the meson dominant HG pressure is always less than the QGP pressure.. This region can be interpreted in terms of the dual description of mostly the quarks and gluons together with $\pi$ mesons. The fundamental assumption in our model is that the baryons in the HG possess a hard-core size and there exists a repulsive interaction between two baryons [13]. However, mesons are not subjected to any such force because they do not have any hard-core size. In constructing a first order phase transition it is essential to include the excluded-volume corrections for baryons in the HG and the EOS for QGP phase should also include QCD interaction terms [16]. Mesons at high temperature can fuse into one another but baryons retain their space. So at large $\mu_{B}$, the fractional occupied volume R is finite and hence mobility of baryons is affected. Therefore, for any low temperature T, we find a corresponding $\mu_{B}$ at which the QGP pressure becomes equal to the HG pressure and beyond which the QGP pressure dominates. At higher T also, this continues unless we reach the end point at which the QGP pressure is always larger than the HG pressure. This is defined as the critical end point in our model. The physical mechanism in this calculation is analogous to the percolation model [14] where a first order transition is obtained through 'jamming' of baryons without any comparison to the QGP. Recent progress and results are encouraging in this direction, but much more work still needs to be done further before this picture becomes conclusive.

CPS and PKS acknowledge the financial support from Department of Science and Tecnology (DST), New Delhi and SKT is grateful to CSIR, New Delhi for the financial support.

\end{document}